\documentclass[12pt,twoside,frenchb,english]{article}
\usepackage{bookman}
\usepackage[T1]{fontenc}
\usepackage[latin1]{inputenc}
\usepackage{amsmath}
\usepackage{graphicx}
\usepackage{amssymb}

\makeatletter

\providecommand{\LyX}{L\kern-.1667em\lower.25em\hbox{Y}\kern-.125emX\@}

\usepackage{bookman}
\usepackage{a4}
\usepackage{cite}
\input{epsfig.sty}
\def\fnum@table{\tablename~{\bf\thetable}}
\def\fnum@figure{\figurename~{\bf\thefigure}}
\def\tablename{\footnotesize{\bf Table}}
\def\figurename{\footnotesize{\bf Figure}}

\AtBeginDocument{

}

\usepackage{babel}
\makeatother
\begin{document}
\begin{center}~\end{center}

\vspace{-0.5cm}
\begin{center}\textbf{\huge Ultra-High Energy Cosmic Rays:} {\huge Some}
\textbf{\huge }{\huge General Features, and Recent Developments Concerning
Air Shower Computations}\end{center}{\huge \par}

\begin{center}\textbf{\large K. Werner$^{a}$ , V. Chernatkin$^{a}$,
R. Engel$^{b}$, N.N. Kalmykov$^{c}$, S. Ostapchenko$^{d}$, T. Pierog$^{b}$}\end{center}{\large \par}

\selectlanguage{frenchb}
\begin{quote}
\textit{$^{a}$SUBATECH, Université de Nantes -- IN2P3/CNRS -- EMN,
 Nantes, France}

\textit{$^{b}$FZK, Institut für Kernphysik, Karlsruhe, Germany}

$^{c}$Skobeltsyn Inst. of Nuclear Physics, Moscow State University,
Russia

$^{d}$\emph{Institut f. Experimentelle Kernphysik, University of
Karlsruhe, Germany}\\

\end{quote}
\begin{center}\emph{Invited talk given at the} \foreignlanguage{english}{\emph{Hadron-RANP2004
workshop}}\\
\end{center}

\begin{quote}
We present an introductory lecture \foreignlanguage{english}{on general
features} of \foreignlanguage{english}{\textbf{}cosmic rays}, for
non-experts, and some recent developments concerning cascade equations
for air shower developments.
\end{quote}

\selectlanguage{english}
\section{Discovery}

The history of cosmic rays goes back to Coulomb, who realized that
a suspended charged metal sphere looses charge. This fact has been
unexplained for a long time, until CTR Wilson (1900) found that this
is due to the ionization of the surrounding air.

In order to learn about the origin of this ionization, Wulf (1910)
climbed the Eifel tower and found that the ionization decreases with
increasing altitude, so the ionization seemed to be linked to ground.

To further investigate this question, Victor Hess (1912) made a balloon
flight up to an altitude of 5$\, $km. He realized that the ionization
increases again above 1$\, $km, which means that the radiation is
coming from above. These findings are generally considered to be the
discovery of cosmic rays.

Pierre Auger (1932) finally detected coincidences between counters
up to 300$\, $m apart, he indentified in this way the first extended
air showers (EAS).

Today we know: When a high energy proton or nucleus hits an air molecule,
it creates many hadrons, which make further collisions, creating more
particles, and so on, leading to a so-called hadronic cascade.

Certain hadrons (mainly $\pi ^{0}$) decay into photons, each of them
initiating an electromagnetic cascade (of photons and electrons).

\section{Characteristics of Cosmic Rays}

\begin{figure}[htb]
\begin{center}\includegraphics[  bb=51bp 107bp 514bp 786bp,
  scale=0.5,
  angle=270,
  origin=c]{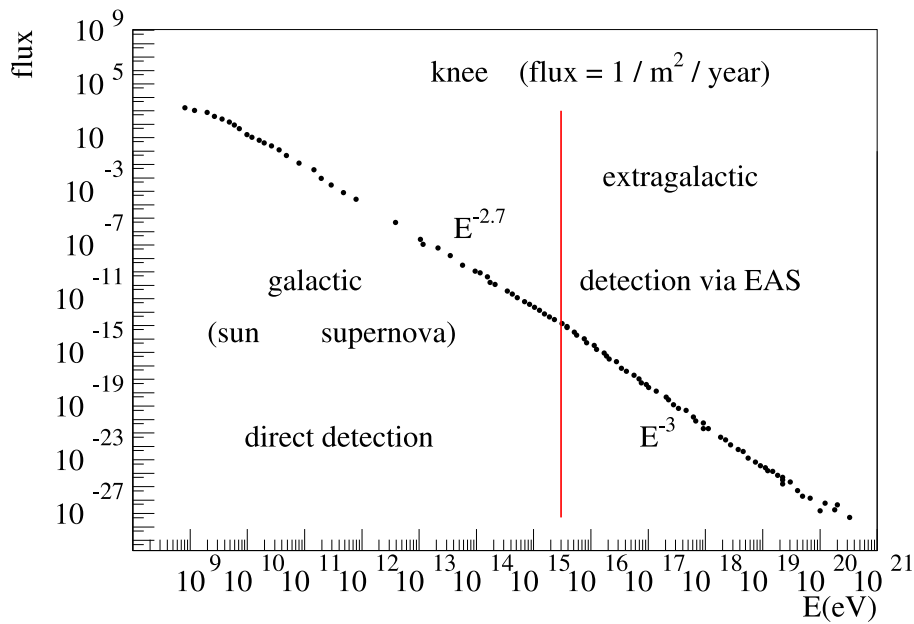}\vspace{-2cm}\end{center}

\caption{The spectrum of cosmic rays\label{cap:The-spectrum}}
\end{figure}

In fig. \ref{cap:The-spectrum}, we show the spectrum of cosmic rays
as measured over the years by many experiments. One observes a surprisingly
simple structure, the spectrum is essentially a power law $E^{-\gamma }$,
with an exponent $2.7$ below $3\, 10^{15}$eV and $3$ above. The
position where the slope changes is usually referred to as the knee.

In particular interesting is the high energy end of the spectrum:
Protons faster than $10^{20}\, $eV interact with the cosmic microwave
background (making $\Delta $ resonances), so they cannot come from
far away; one expects therefore a galactic source. On the other hand
the Larmor radius suggests an extragalactic source... .

\section{Sources}

What are the sources of the high energy cosmic rays? Two scenarios
are considered \cite{soucre1,source2}:

\begin{itemize}
\item Top-Down scenario, which amounts to the decay of some extremely heavy
particle
\item Bottom-Up scenario, where particles are accelerated by conventional
electromagnetic forces. Electric fields are difficult to find, so
probably one has to deal with magnetic fields (Fermi)
\end{itemize}
Fermi proposed that particles may be accelerated by entering and leaving
moving magnetic clouds (like tennis balls), which leads to a gain
of energy of $\Delta E/E=4/3\: \beta ^{2}$ (second order), or particles
are accelerated by entering and leaving moving shock fronts, with
$\Delta E/E=4/3\: \beta $ (first order).

In any case, we have $\Delta E=\alpha E$. Considering $n$ iterations,
one finds \[
E=E_{0}(1+\alpha )^{n}.\]
Supposing an escape probability $\varepsilon $, we find for the number
of particles\[
N=N_{0}(1-\varepsilon )^{n}.\]
So we get indeed an exponential spectrum,\[
\frac{N}{N_{0}}=\left(\frac{E}{E_{0}}\right)^{\ln (1-\varepsilon )/\ln (1+\alpha )}.\]

Possible accelerators are 

\begin{itemize}
\item Supernova shock waves $\leq 10^{15}\, $eV.
\item Jets in AGNs: the jets create shocks into the intergalactic medium.
\item Rotating neutron stars (ms spin rates, $B=10^{13}\, $G): Dynamo effect
$\to $$10^{20}\, $V.
\end{itemize}

\section{Air Showers}

The first step is a primary hadronic interactions of a proton or a
nucleus with air producing many hadrons. Then these secondary hadrons
interact with air, giving rise again to hadron production, and so
on. This is the so-called hadronic cascade. Hadrons may decay before
suffering an interaction, for example:\begin{eqnarray*}
\pi ^{0} & \to  & \gamma \gamma \quad (98.8\%)\\
 &  & \\
\pi ^{\pm } & \to  & \mu ^{\pm }+\nu (\bar{\nu })\quad (100\%)\\
 &  & \\
K^{\pm } & \to  & \mu ^{\pm }+\nu (\bar{\nu })\quad (63\%).
\end{eqnarray*}
Most muons reach the ground (at high energies). 

Each $\gamma $ initiates an electromagnetic cascade (photons and
electrons). In an EM cascade, photons interact via

\begin{itemize}
\item Compton scattering with electrons from an atom (low $E$),
\item Photoelectric effect: the photon is absorbed by an atom, its energy
is used to liberate an electron (low $E$).
\item Pair production: the photon interacts with the Coulomb field of a
nucleus to produce a electron-positron pair (high $E$).
\end{itemize}
Electrons interact via

\begin{itemize}
\item Ionization energy loss.
\item Bremsstrahlung: the electron interacts with the Coulomb field of a
nucleus to produce a slowed down electron and plus the radiated photon.
\end{itemize}
At \textbf{}high energy, the most important processes are pair production
and Bremsstrahlung: \begin{eqnarray*}
e & \to  & e+\gamma \\
 &  & \\
\gamma  & \to  & e+e,
\end{eqnarray*}
so in any case the number of particles doubles for each iteration.

The splitting continues till some critical energy $E_{c}$ is reached
(after this the particles get absorbed or decay only)

We first discuss a toy model \textbf{}(Heitler 1944):

\begin{itemize}
\item At each step the energy is given in equal parts to the two daughters.
\item A branching happens always after a collision length $\lambda $, so
the number of branchings at depth $X$ is $X/\lambda $.
\end{itemize}
The number of particles and the energy per particle are\[
N(X)=2^{X/\lambda },\quad E(X)=E_{0}/N.\]
The shower maximum is at $E=E_{c}$, so \[
N(X_{\max })=2^{X_{\max }/\lambda }=\frac{E_{0}}{E_{c}}\: \to \: X_{\max }=\lambda \frac{\ln E_{0}/E_{c}}{\ln 2}.\]
So we find the important \underbar{general results}:\begin{eqnarray*}
N_{\max } & \propto  & E_{0}\, ,\\
X_{\max } & \propto  & \ln E_{0}\, .
\end{eqnarray*}

The standard technique to calculate air shower developments is the
Monte Carlo technique \cite{mc}: Each particle may decay or interact,
whatever happens earlier, at the corresponding depth; new particles
from decay or interaction are written to stack and treated subsequently.

This technique is easy to implement, but \textbf{}very slow at high
energies, since $N\propto E_{0}$ and even impossible at $10^{20}$eV.

An alternative is the numerical solution of so-called cascade equations
(CE) \cite{ce1,ce2,ce3,ce4}, which is at high energies considerably
faster than the MC approach.

\section{Characteristics of Air Showers}

In the following we show results from the CE calculations (=Conex)
for proton induced showers (if not mentioned otherwise), and some
comparisons with the CORSIKA Monte Carlo calculations \cite{mc}.
The cascade equations will be discussed in the following chapters.
The final aim being the solution of the full threedimensional cascade
equations, we show here only results concerning the $X$ (depth) dependence
of particle numbers, based on the $1D$ version of the approach, which
makes use of the small angle approximation. There is no information
about the lateral distributions.

\begin{figure}[htb]
\begin{center}~~\\
\includegraphics[  bb=56bp 84bp 530bp 768bp,
  scale=0.35,
  angle=270,
  origin=c]{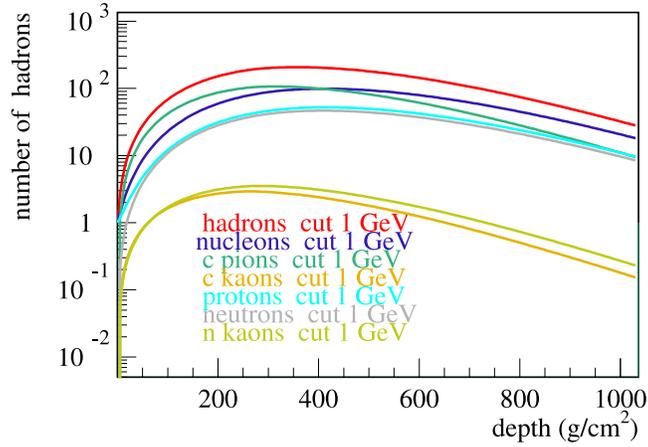}\vspace{-2cm}\end{center}

\caption{Produced hadrons at 10\textasciicircum{}14 eV.}

~~
\end{figure}

\begin{figure}[htb]
\begin{center}\includegraphics[  scale=0.35,
  angle=270,
  origin=c]{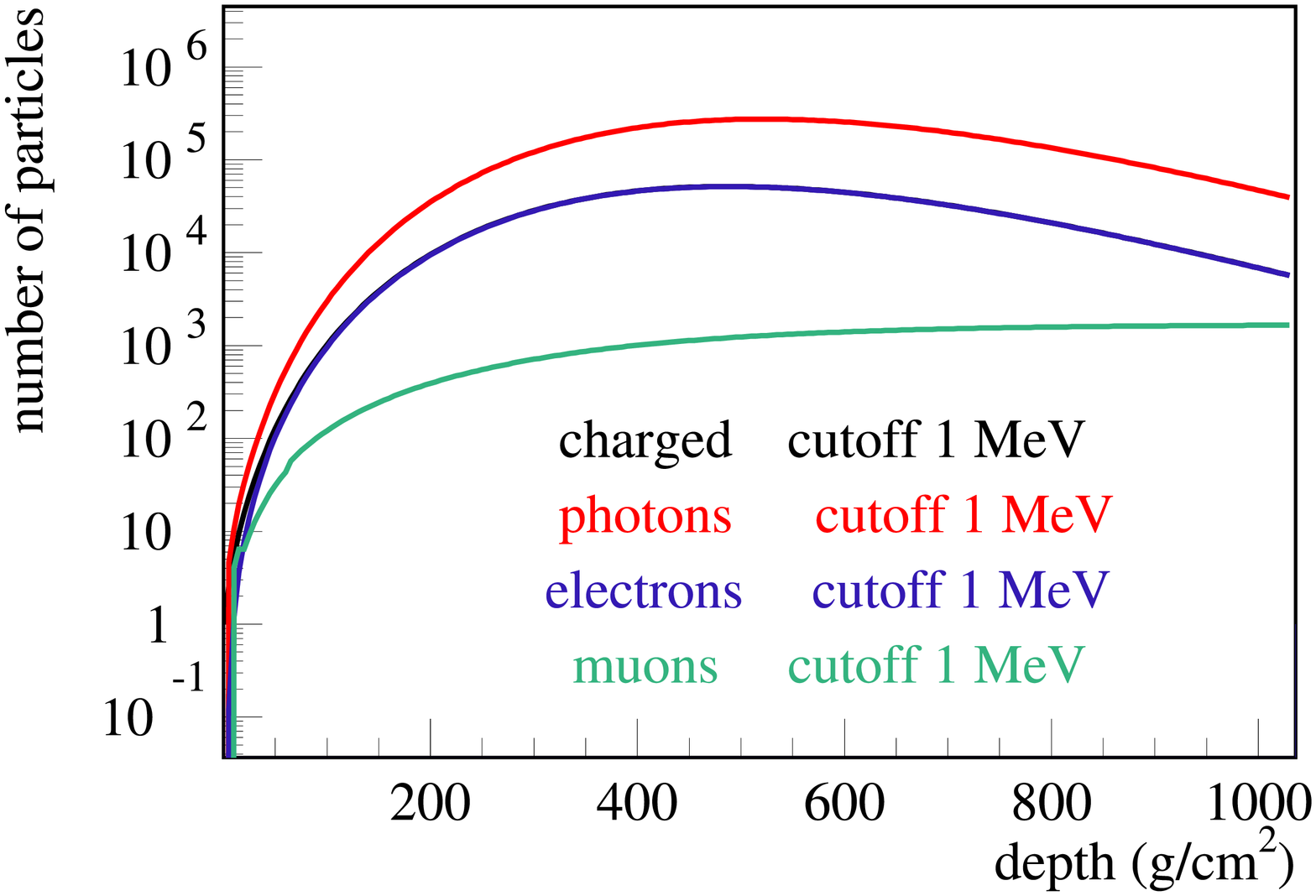} \vspace{-1cm}\end{center}

\caption{Produced leptons, photons at 10\textasciicircum{}14 eV.}
\end{figure}

\begin{center}%
\begin{figure}[htb]
\begin{center}\includegraphics[  scale=0.35,
  angle=270,
  origin=c]{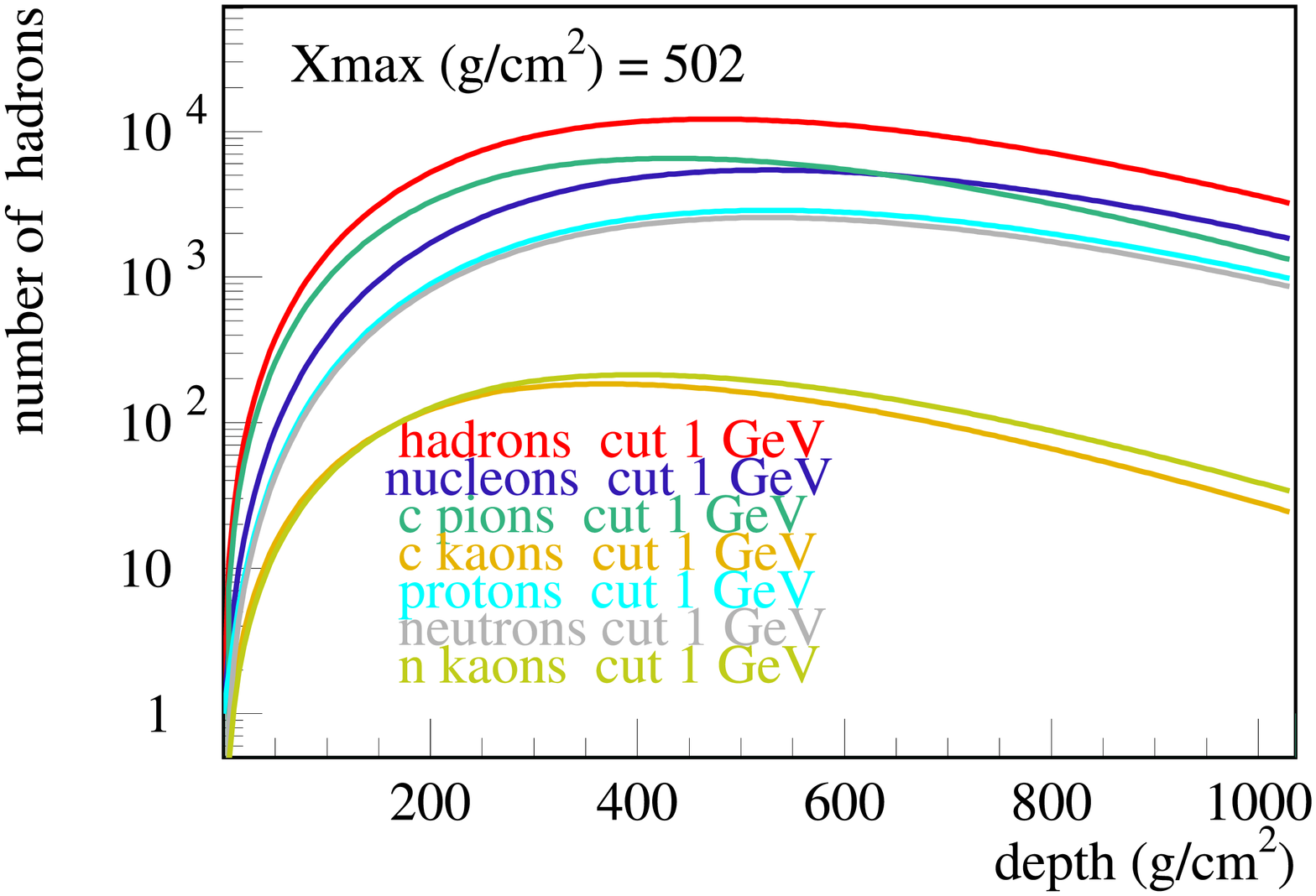}\vspace{-1cm}\end{center}

\caption{Produced hadrons at 10\textasciicircum{}16 eV  }
\end{figure}
\end{center}

\begin{figure}[htb]
\begin{center}\includegraphics[  scale=0.35,
  angle=270,
  origin=c]{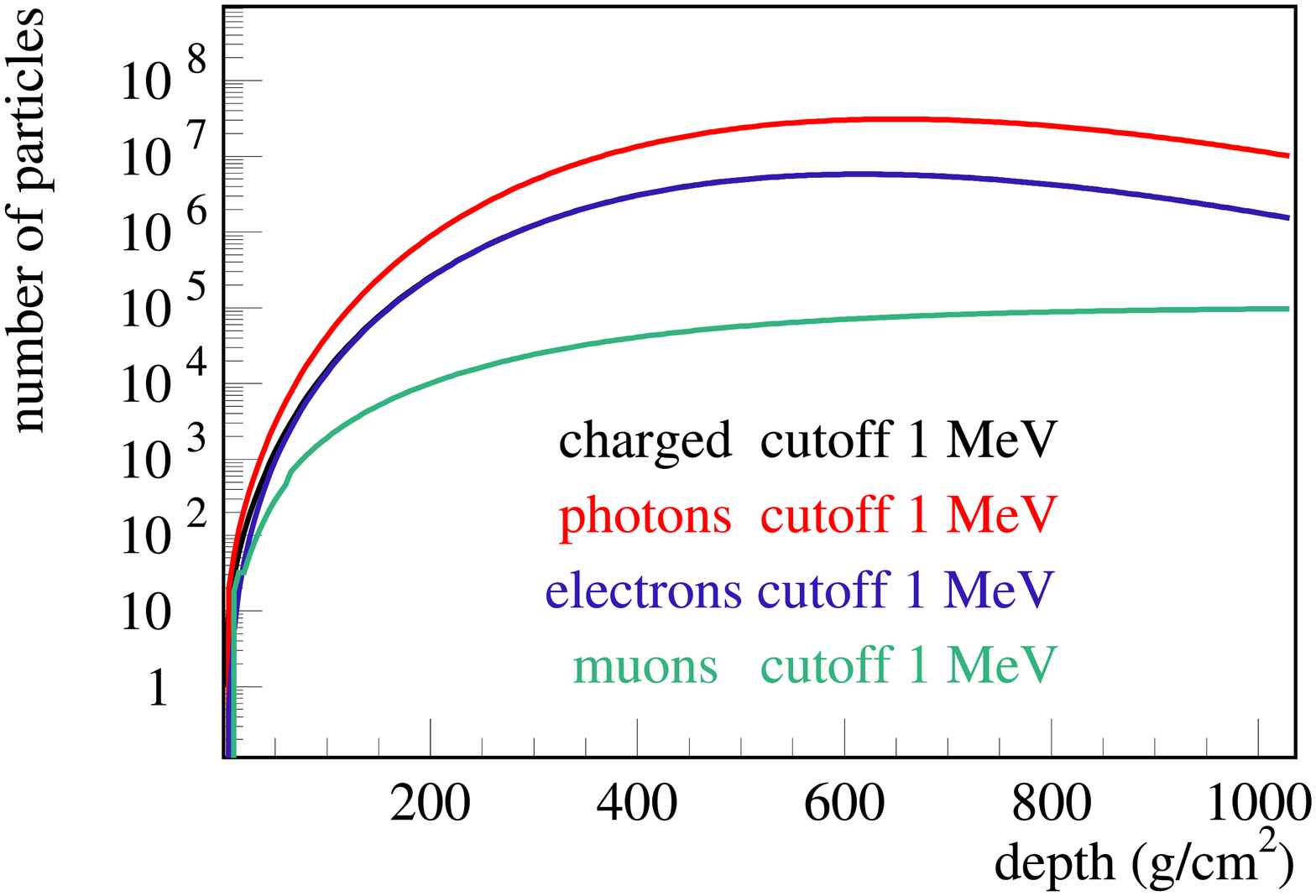}\vspace{-2cm}\end{center}

\caption{Produced leptons, photons at 10\textasciicircum{}16 eV  }

~~~
\end{figure}

\begin{center}%
\begin{figure}[htb]
\begin{center}\includegraphics[  scale=0.35,
  angle=270,
  origin=c]{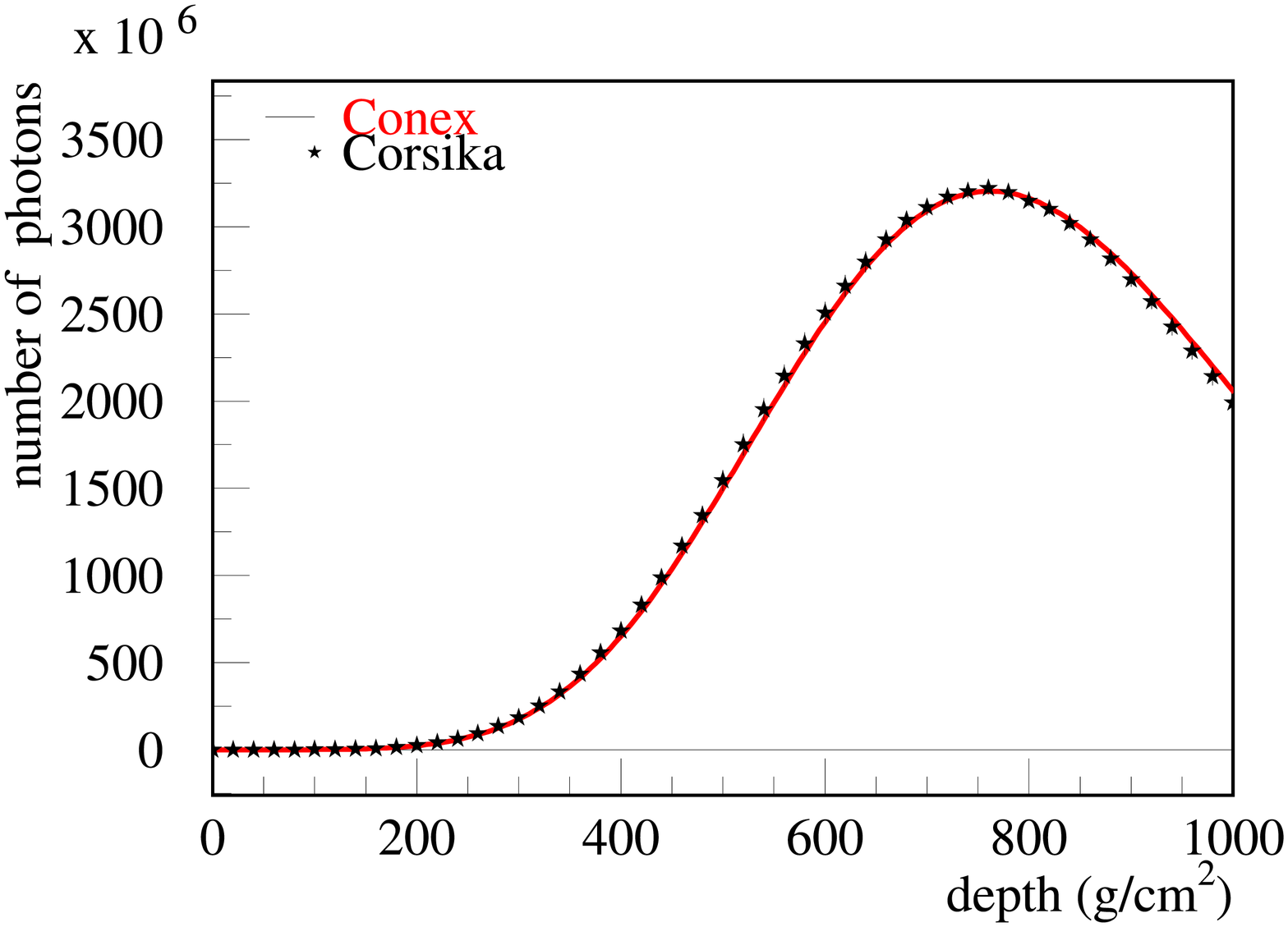}\vspace{-1cm}\end{center}

\caption{Comparison Corsika-Conex at 10\textasciicircum{}18 eV.}
\end{figure}
\end{center}

\begin{figure}[htb]
\begin{center}\includegraphics[  scale=0.35,
  angle=270,
  origin=c]{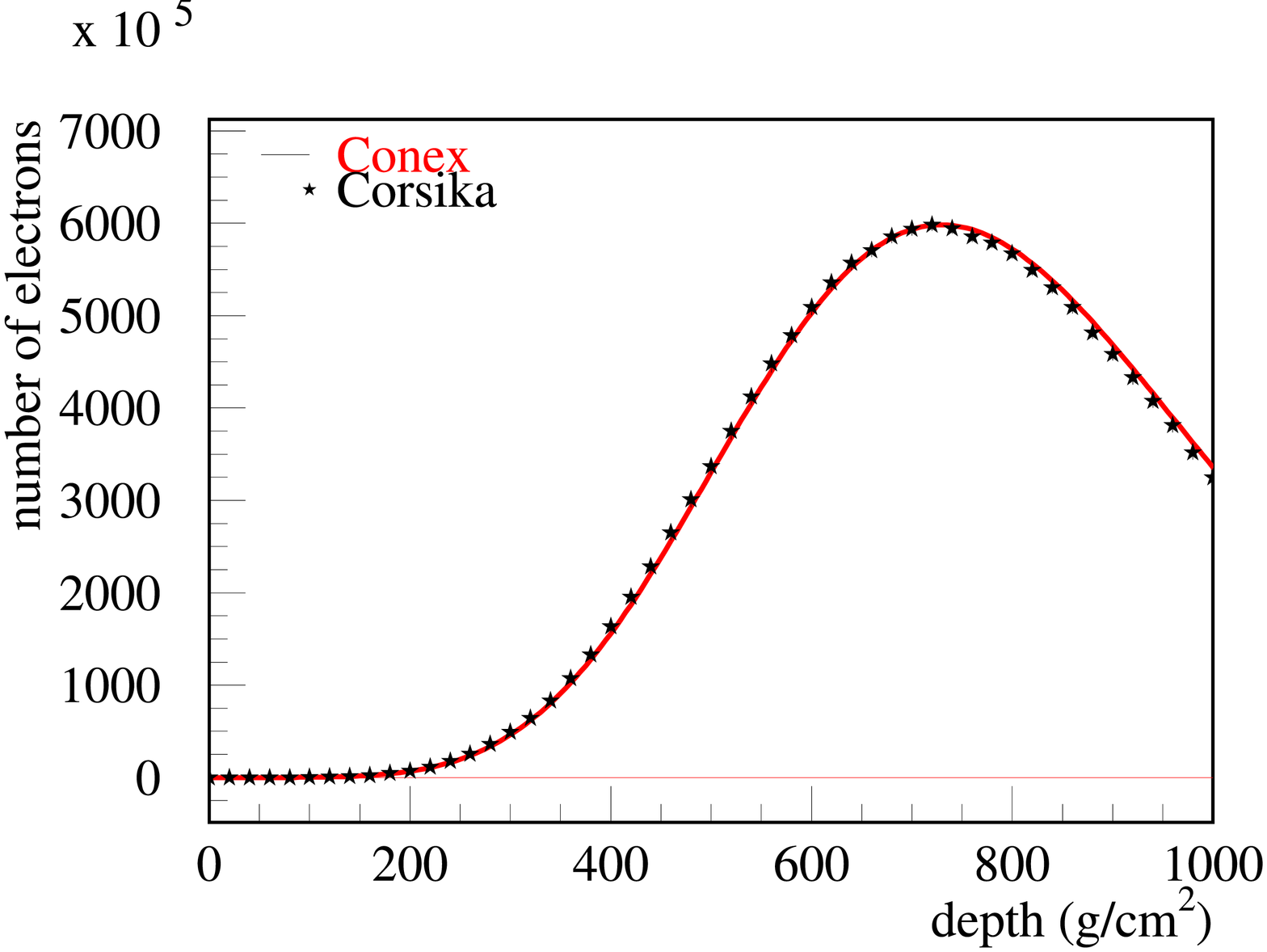}\vspace{-1cm}\end{center}
\vspace{-1cm}

\caption{Comparison Corsika-Conex at 10\textasciicircum{}18 eV.}
\end{figure}

\begin{figure}[htb]
\begin{center}\includegraphics[  scale=0.35,
  angle=270,
  origin=c]{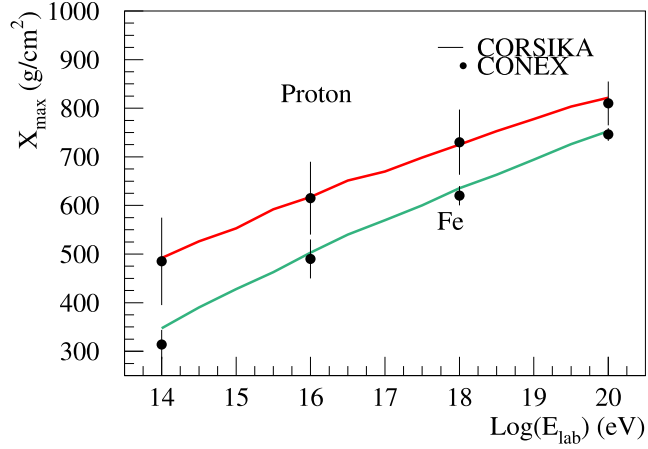}\vspace{-1cm}\end{center}

\caption{Shower maximum .  }
\end{figure}

\begin{figure}[htb]
\begin{center}\includegraphics[  scale=0.35,
  angle=270,
  origin=c]{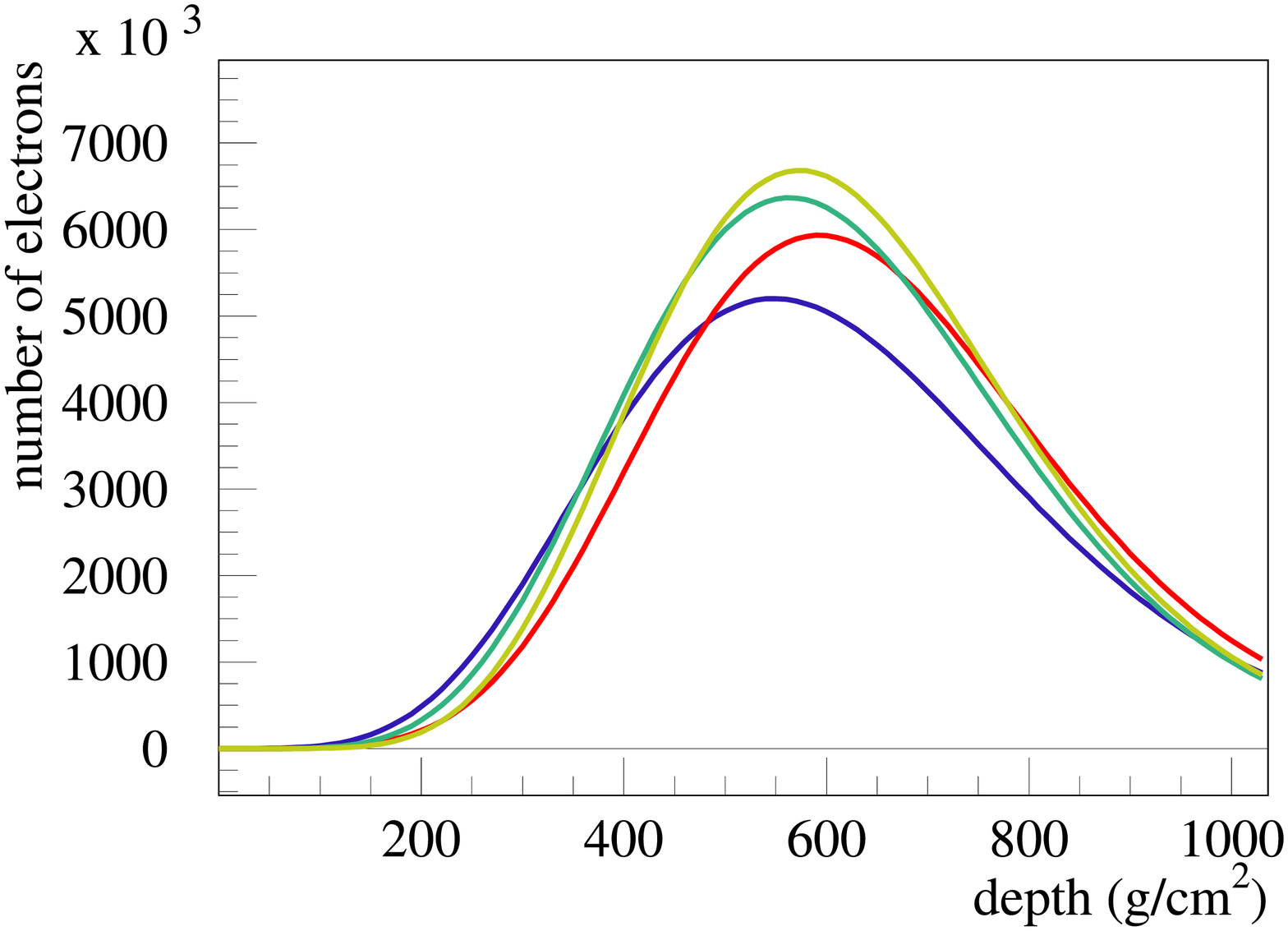}\vspace{-1cm}\end{center}

\caption{Individual showers at 10\textasciicircum{}16 eV  . }
\end{figure}

~

~

~

~

\section{Hadronic 1D cascade equations }

Here we discuss the equations for the energy spectra $h_{n}(E,X)$
of hadrons of type $n$ at depth $X$, where the depth is defined
as \[
X=\int _{P}^{\infty }\, \rho (x)dx,\]
with $\rho $ being the density of air. The decrease of $h_{n}(E,X)$
due to collisions is\[
\frac{dh_{n}}{dX}=-\frac{h_{n}}{\lambda _{n}},\]
where the mean inelastic free path $\lambda _{n}$ is \[
\lambda _{n}=\frac{m_{\mathrm{air}}}{\sigma _{\mathrm{inel}}^{(n)}}.\]
The decay rate in the particle c.m. system is $dh_{n}/d\tau =-h_{n}/\tau _{n}$,
so\[
\frac{dh_{n}}{dX}=-\frac{m_{n}\rho ^{-1}(X)}{c\tau _{n}E}\, h_{n}.\]
The equation for hadron density, considering collisions and decay
(only loss terms) is

\[
\frac{\partial h_{n}(E,X)}{\partial X}=-h_{n}(E,X)\left(\frac{1}{\lambda _{n}}+\frac{m_{n}\rho ^{-1}(X)}{c\tau _{n}E}\right).\]
The full equation, including gain from higher energies, and a source
term (generalizing simple initial conditions) is \begin{eqnarray*}
\frac{\partial h_{n}(E,X)}{\partial X} & = & -h_{n}(E,X)\left[\frac{1}{\lambda _{n}(E)}+\frac{m_{n}\rho ^{-1}(X)}{c\tau _{n}E}\right]\\
 &  & +\sum _{m}\int _{E}^{E_{\mathrm{max}}}h_{m}(E',X)\\
 &  & \qquad \left[\frac{W_{mn}(E',E)}{\lambda _{m}(E')}+D_{mn}(E',E)\: \frac{m_{m}\rho ^{-1}(X)}{c\tau _{m}E'}\right]dE'\\
 &  & +S_{n}(E,X).
\end{eqnarray*}
where $W_{mn}(E',E)$ and $D_{mn}(E',E)$ are inclusive spectra of
secondaries of type $n$ and energy$\, E$ produced in interactions
($W$) or decays ($D$) of primaries of type $m$ and energy $E'$. 

To obtain a system of differential equations, one proceeds to an energy
discretization as\[
E_{i}=E_{\mathrm{min}}\cdot c^{i},\]
which gives \[
\frac{\partial h_{ni}(X)}{\partial X}=-h_{ni}(X)\left[\frac{1}{\lambda _{ni}}+\frac{m_{n}\rho ^{-1}(X)}{c\tau _{n}E_{i}}\right]+\sum _{m}\sum _{j=i}^{j_{\mathrm{max}}}h_{mj}(X)\left[\frac{W_{mn}^{ji}}{\lambda _{mj}}+D_{mn}^{ji}\frac{m_{m}\rho ^{-1}(X)}{c\tau _{m}E_{j}}\right]+S_{ni}(X),\]
 with $h_{ni}(X)=$$h_{n}(E_{i},X)$, $\lambda _{ni}=\lambda _{n}(E_{i})$,
and\begin{eqnarray*}
W_{mn}^{ji} & = & \int _{E_{i-1}}^{E_{i}}f(\frac{E}{E_{i}})W_{mn}(E_{j},E)dE+\int _{E_{i}}^{E_{i+1}}g(\frac{E}{E_{i+1}})W_{mn}(E_{j},E)dE,\\
D_{mn}^{ji} & = & \int _{E_{i-1}}^{E_{i}}f(\frac{E}{E_{i}})D_{mn}(E_{j},E)dE+\int _{E_{i}}^{E_{i+1}}g(\frac{E}{E_{i}})D_{mn}(E_{j},E)dE,
\end{eqnarray*}
with\[
f(x)=\frac{cx-1}{c-1},\quad g(x)=1-f(x),\]
in order to assure energy and particle number conservation.

The solution of the homogeneous equation is\[
h_{ni}(X)=h_{ni}(X_{o})\, \exp \left\{ -\frac{(1-W_{nn}^{ii})\, (X-X_{o})}{\lambda _{ni}}-\frac{m_{n}\, (L(X)-L(X_{o}))}{c\tau _{n}E_{i}}\right\} ,\]
with $L'=\rho ^{-1}$. The solution of the full equation is

\begin{eqnarray*}
h_{ni}(X) & = & h_{ni}(X_{o})\, \exp \left\{ -\frac{(1-W_{nn}^{ii})\, (X-X_{o})}{\lambda _{ni}}-\frac{m_{n}\, (L(X)-L(X_{o}))}{c\tau _{n}E_{i}}\right\} \\
 &  & \\
 &  & +\int _{X_{o}}^{X}\left\{ \sum _{m}\sum _{j=i+1}^{j_{max}}h_{mj}(X')\left[\frac{W_{mn}^{ji}}{\lambda _{mj}}+D_{mn}^{ji}\frac{m_{m}\rho ^{-1}(X')}{c\tau _{m}E_{j}}\right]+S_{ni}(X')\right\} \\
 &  & \\
 &  & \qquad \exp \left\{ -\frac{(1-W_{nn}^{ii})\, (X-X')}{\lambda _{ni}}-\frac{m_{n}\, (L(X)-L(X'))}{c\tau _{n}E_{i}}\right\} dX'
\end{eqnarray*}

\section{Electromagnetic cascade equations}

We take the longitudinal coordinate to be $z$, the lateral coordinates
$x$ and $y$. The direction of a particle at a given point $(x,y,z)$
may be written as\[
\vec{u}=\left(\begin{array}{c}
 \sin \theta \cos \phi \\
 \sin \theta \sin \phi \\
 \cos \theta \end{array}
\right).\]
 We define\[
\alpha =\sin \theta \cos \phi \, ,\]
 \[
\beta =\sin \theta \sin \phi \, .\]
In the following we work with the coordinates $\alpha $, $\beta $,
$x$, $y$, $z$ of a particle and its energy $E$. We consider differential
spectra of electrons, positrons, and gammas\[
l_{n}(\alpha ,\beta ,x,y,z,E)=\frac{dN_{n}}{d\alpha \, d\beta \, dx\, dy\, dE}\, ,\]
with $n\in \{e^{+},\, e^{-},\, \gamma \}$, for a given depth $z$.
The energy $E$ refers to the kinetic energy of electrons and total
energy of photons. We suppose a source term $S_{n}=S_{n}(\alpha ,\beta ,x,y,z,E)$.
In the following we will supress the arguments of the fuctions $l_{n}$,
$S_{n}$ etc, unless explicitely needed. The equations of electron-photon
cascades may be written in the following form \[
\vec{u}\, \vec{\nabla }\, l_{n}=L_{n}[l]+L_{\mathrm{Coul},n}[l_{n}]+S_{n},\]
with\[
\vec{u}\, \vec{\nabla }=\! \cos \theta \frac{\partial }{\partial z}\! +\! \alpha \frac{\partial }{dx}\! +\! \beta \frac{\partial }{\partial y}.\]
and\[
L_{n}[l]=\sum _{m}\int _{E}^{E_{0}}l_{m}(E')\, W_{mn}(E',E)\, dE',\]
 with the differential spectra\begin{eqnarray*}
W_{e^{-}e^{-}}(E',E) & = & W_{\mathrm{brems}}(E',E'-E)+W_{\mathrm{Moeller}}(E',E)\, \Theta (E-E_{\mathrm{cut}})-(\sigma _{e^{-}}(E)-\frac{\overleftarrow{d}}{dE}\beta )\, \delta (E'-E),\\
W_{e^{+}e^{+}}(E',E) & = & W_{\mathrm{brems}}(E',E'-E)+W_{\mathrm{Bhabha}}(E',E)\, \Theta (E-E_{\mathrm{cut}})-(\sigma _{e^{+}}(E)-\frac{\overleftarrow{d}}{dE}\beta )\, \delta (E'-E),\\
W_{e^{+}e^{-}}(E',E) & = & W_{\mathrm{Bhabha}}(E',E)\, \Theta (E-E_{\mathrm{cut}}),\\
W_{\gamma \, e^{-}}(E',E)\,  & = & W_{\mathrm{pair}}(E',E)+W_{\mathrm{Compton}}(E',E'-E),\\
W_{\gamma \, e^{+}}(E',E)\,  & = & W_{\mathrm{pair}}(E',E),\\
W_{e^{-}\gamma }(E',E)\,  & = & W_{\mathrm{brems}}(E',E),\\
W_{e^{+}\gamma }(E',E)\,  & = & W_{\mathrm{brems}}(E',E)+W_{\mathrm{annih}}(E',E),\\
W_{\gamma \, \gamma }(E',E)\;  & = & W_{\mathrm{Compton}}(E',E)-\sigma _{\gamma }(E)\, \delta (E'-E),
\end{eqnarray*}
 and\[
\sigma _{n}(E)=\sum _{m}\int _{0}^{E}W_{nm}(E,E')\, dE'.\]
The ionization energy loss coefficient is obviously zero for photons,
$\beta _{\gamma }=0$, the same as the Coulomb term, $L_{\mathrm{Coul},\gamma }=0$.
Otherwise\[
L_{\mathrm{Coul},e}[l_{e}]=\int d\eta d\zeta \left\{ l_{e}(\alpha +\eta ,\beta +\zeta ,...)-l_{e}(\alpha ,\beta ,...)\right\} \frac{d\sigma _{\mathrm{Coul}}(\eta ,\zeta )}{d\eta d\zeta },\]
where $l_{e}$ stands for $l_{e^{-}}$ or $l_{e^{+}}$. An expansion
gives\[
l_{e}(\alpha +\eta ,\beta +\zeta ,...)-l_{e}(\alpha ,\beta ,...)=\sum _{\begin{array}{c}
 i,j\\
 i+j>0\end{array}
}\frac{\partial ^{i+j}l_{e}}{\partial \alpha ^{i}\partial \beta ^{j}}\, \frac{\eta ^{i}\zeta ^{j}}{i!j!}.\]
So\[
L_{\mathrm{Coul},n}[l_{n}]=\sum _{\begin{array}{c}
 i,j\\
 i+j>0\end{array}
}C_{ij}^{n}\, \cos ^{2i}\! \theta \, \frac{\partial ^{2i+2j}l_{e}}{\partial \alpha ^{2i}\partial \beta ^{2j}}.\]
with\begin{equation}
C_{ij}^{\gamma }=0,\qquad C_{ij}^{e}=\frac{1}{2i!\, 2j!}\int d\vartheta ^{2}d\varphi \, \frac{\eta ^{2i}\zeta ^{2j}}{\cos ^{2i}\! \theta }\, \frac{1}{2\pi }\, \frac{d\sigma _{\mathrm{Coul}}(\vartheta ^{2})}{d\vartheta ^{2}},\end{equation}
So we finally have

\[
(\cos \theta \frac{\partial }{\partial z}\! +\! \alpha \frac{\partial }{dx}\! +\! \beta \frac{\partial }{\partial y})l_{n}=L_{n}[l]+\! \! \! \! \sum _{\begin{array}{c}
 ij\\
 i+j>0\end{array}
}\! \! \! \! C_{ij}^{n}\, \cos ^{2i}\! \theta \, \frac{\partial ^{2i+2j}l_{n}}{\partial \alpha ^{2i}\partial \beta ^{2j}}+S_{n}.\]
In the small angle approximation one may integrate over the variables
$\alpha ,\beta ,x,y$ and one obtaines for \[
\tilde{l}_{n}(z,E)=\int l(\alpha ,\beta ,x,y,z,E)d\alpha d\beta dxdy\]
the equation\[
\frac{\partial }{\partial z}\tilde{l}_{n}=L_{n}[\tilde{l}]+S_{n},\]
which can be solved after energy discretisation without major problems.

\section{Summary (Cascade Equations)}

CONEX 1D works already quite well, it gives practically the same results
as the Monte Carlo calculations with Corsika, although the time needed
for the computations is enormously reduced. Concerning the 3D version,
there is work in progress.

\end{document}